\documentclass[reprint,
superscriptaddress,
 amsmath,amssymb,
 aps,
]{revtex4-2}

\usepackage{graphicx}
\usepackage{dcolumn}
\usepackage{bm}
\usepackage[colorlinks=true, linkcolor=blue, urlcolor=blue,citecolor=blue]{hyperref}

\usepackage{xcolor}
\usepackage{amssymb}
\usepackage{pifont}
\begin{document}


\title{Quantum geometric origins of the orbital degrees of freedom of hybrid bosonic quasiparticles in magnetic systems}

\author{D. Quang To}%
\email{quangto@udel.edu}
\affiliation{Department of Materials Science and Engineering, University of Delaware, Newark, Delaware 19716, USA}

\author{Dai Q. Ho}
\affiliation{Department of Materials Science and Engineering, University of Delaware, Newark, Delaware 19716, USA}
\affiliation{Faculty of Natural Sciences, Quy Nhon University, Quy Nhon 55113, Vietnam}

\author{Joshua M. O. Zide}%
\affiliation{Department of Materials Science and Engineering, University of Delaware, Newark, Delaware 19716, USA}%

\author{Lars Gundlach}
\affiliation{Department of Chemistry and Biochemistry, University of Delaware, Newark, Delaware 19716, USA}

\author{M. Benjamin Jungfleisch}
\affiliation{Department of Physics and Astronomy, University of Delaware, Newark, Delaware 19716, USA}

\author{Garnett W. Bryant}
\affiliation{Nanoscale Device Characterization Division, Joint Quantum Institute, National Institute of Standards and Technology, Gaithersburg, Maryland 20899-8423, United States}
\affiliation{University of Maryland, College Park, Maryland 20742, USA}

\author{Anderson Janotti}
\affiliation{Department of Materials Science and Engineering, University of Delaware, Newark, Delaware 19716, USA}

\author{Matthew F. Doty}%
 \email{doty@udel.edu}
\affiliation{Department of Materials Science and Engineering, University of Delaware, Newark, Delaware 19716, USA}

\date{\today}

\begin{abstract}
The orbital degree of freedom has recently attracted significant attention due to the novel phenomena it enables in condensed matter systems. However, the interpretation of the orbital degree of freedom in bosonic quasiparticles remains conceptually ambiguous and the mechanisms governing the transfer of orbital angular moment (OAM) between distinct quasiparticles, such as magnons and phonons, are not yet fully understood. We investigate orbital dynamics in bosonic systems and identify two origins of OAM: (i) global rotational motion of the system, and (ii) the quantum geometry of wavefunctions. Focusing on the latter, we study strongly coupled magnon–phonon systems in two-dimensional antiferromagnets as a test case. We uncover finite OAM arising from quantum geometric effects via two mechanisms: (a) time-parity symmetry breaking, yielding intra-band OAM, and (b) interband coupling, generating inter-band OAM. We propose that an electrical detection scheme based on the transverse voltage generated by hybrid magnon–phonon modes can be used to experimentally probe the bosonic orbital degree of freedom. Our results establish a foundation for the emerging field of phonon orbitronics, providing both a conceptual bridge between phonon and magnon orbitronics and a tool for better understanding magnon-polarons. They also advance a unified framework for harnessing orbital degrees of freedom in bosonic systems and pave the way toward electrical control of magnetization and phononic transport.
\end{abstract}

\maketitle
\textit{Introduction} -- The properties of quantum materials are governed by a rich interplay between particles and quasiparticles such as photons, electrons, phonons, magnons, or excitons, each of which carries distinct degrees of freedom such as charge, mass, spin, orbital, or valley index. When particles and quasiparticles strongly interact, their coupled degrees of freedom can give rise to novel and intriguing emergent phenomena \cite{Keimer2017,Tokura2017,Basov2017,Tabuchi2015,Bhoi2019,Golovchanskiy2021,To2024}. Recently, significant attention has been directed toward the orbital degrees of freedom (ODF) in electrons and magnons. In electronic systems, the orbital angular momentum of electrons underlies phenomena such as orbital magnetization \cite{Thonhauser2005,Shi2007,Resta2010,Thonhauser2011}, the orbital Hall effect \cite{Go2018,Choi2023,Lyalin2023,Gupta2025}, and orbital-to-spin or orbital-to-charge interconversion \cite{Ding2020,Varotto2022,Pezo2024,Nikolaev2024,Bony2025}. In magnetic materials, the orbital moment of magnons plays a crucial role in thermal and spin transport\cite{Neumann2020,Fishman2022,Go2024,To2025b,Neumann2025}. For example, it gives rise to the magnon orbital Seebeck and magnon orbital Nernst effect and contributes to the electric activity of magnons, enabling their detection and manipulation via electric probes \cite{Neumann2024,To2025,Neumann2025}. Recent advancements have established the foundations of the rapidly emerging field of electronic and magnetic orbitronics \cite{Go2021,Das2023,Jo2024}. 

In phononic systems, studies have primarily focused on the pseudo-spin angular momentum (PSAM) \cite{Streib2021}, leading to the definition and exploration of chiral phonons, which are collective lattice vibrations characterized by circular polarization with opposite PSAM \cite{Wang2024,Zhang2014,Zhang2015}. Chiral phonons are responsible for remarkable effects such as giant phonon magnetic moments \cite{Juraschek2019,Juraschek2022,Luo2023,Chaudhary2024}, the chiral-phonon-activated spin Seebeck effect \cite{Kim2023}, valley-selective phonon-electron interactions \cite{Liu2019}, and chiral-phonon-mediated high-temperature superconductivity \cite{Gao2023}. In contrast, investigations into the ODF of phonons remain sparse. Consequently, the role of phonon pseudo orbital angular momentum (POAM) in influencing both phononic properties and interactions with other degrees of freedom, including conversions between phonon POAM and that of other quasiparticles, remains poorly understood. 

While the ODF is a fundamental concept in quantum physics \cite{Griffiths2018}, its definition varies significantly across different types of particles and quasiparticles in condensed matter physics. For example, electrons and atoms that constitute phonons possess real mass, allowing their orbital motions to be interpreted classically and associated with well-defined POAM. In contrast, magnons are massless quasiparticles, rendering classical notions of POAM inapplicable. To avoid this issue, the ODF of magnons is described in terms of the orbital angular moment (OAM), which is directly linked to a measurable quantity, namely electric polarization, rather than to the conventional POAM ~\cite{To2025}. However, the physical interpretation of the ODF of magnons remains unclear, particularly because magnons originate from local spin precession around fixed atomic sites, without any associated real-space motion of atoms. These conceptual ambiguities and discrepancies become especially problematic in hybridized systems such as magnon-polarons where strong coupling between magnons and phonons leads to emergent properties that cannot be understood by treating each quasiparticle independently. In such cases, conventional definitions of, for instance, phonon POAM, which rely on finite mass of atoms that constitute phonons, are not directly applicable and the mechanisms underlying the conversion of ODF between magnons and phonons consequently remain largely unexplored.

In this letter, we address this gap in understanding by investigating the ODF in magnon–phonon coupled systems. We begin by analyzing the ODF of magnon-polarons in bosonic systems and demonstrate that the ODF of bosonic quasiparticles, such as magnons and phonons, originates from two distinct mechanisms: (i) the global rotation of the system and (ii) the quantum geometry of the magnon-polaron wavefunctions, which reflects the internal dynamics of magnon-polaron wavepackets. We formulate the magnon-polaron ODF induced by quantum geometry within a modern theoretical framework that encodes the ODF in the bosonic wavefunction, establishing a formalism that can be applied to both bosonic and fermionic systems. This wavepacket-based contribution to the ODF is linked to the quasi–quantum geometric tensor \cite{Kang2025,Kim2025}, and plays a pivotal role in determining the properties of hybrid bosonic excitations in quantum materials, including magnon–phonon states of magnetic insulators. For example, we find that out-of-plane phonon modes, despite their intrinsic linear polarization along the z-direction, can acquire a finite out-of-plane OAM through hybridization with magnons. We show that this finite OAM, coupled with the magnetic moment of the magnon component, should lead to a measurable transverse voltage when a temperature or strain gradient is applied to excite the magnon-polaron. 

\begin{figure}
\centering
\includegraphics[width=0.5\textwidth]{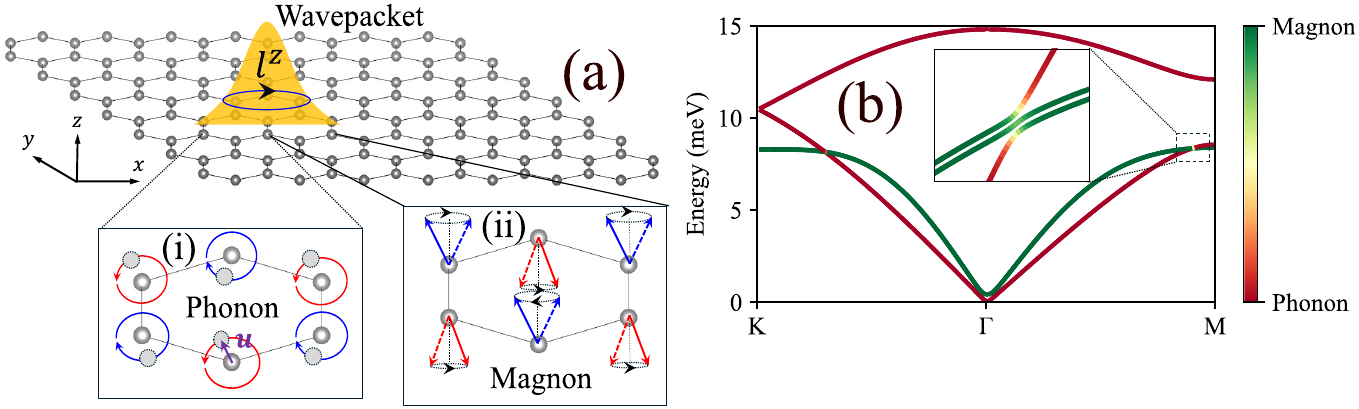}
\caption{(a) Schematic of magnon-polaron wavepacket dynamics in a two-dimensional honeycomb lattice. The wavepacket arises from the hybridization of an elastic wave and a spin wave. Inset (i) shows the elastic wave characterized by the displacement of atoms $\boldsymbol{u}$ around their equilibrium positions and an associated pseudospin due to in-plane circular motion. Inset (ii) shows the spin wave, which carries spin angular momentum originating from the precession of local magnetic moments. The inherent self-rotation of the hybridized wavepacket establishes the orbital degree of freedom, rooted in the quantum geometry of the magnon-polaron Bloch wavefunction. (b) Dispersion along the symmetry path $K-\Gamma-M$ of the phonon (red) and magnon (green) modes of a honeycomb AFM with N\'eel order. Hybridized magnon-polarons emerge at the energy degeneracy point as shown more clearly in the inset.}
\label{Scheme}
\end{figure}

\textit{Pseudo angular momentum of bosonic quasiparticle} -- Unlike electrons, which possess charge, finite mass, intrinsic spin, and orbital degrees of freedom, phonons and magnons are bosonic quasiparticles with fundamentally different properties. Magnons carry intrinsic spin, but lack both mass and charge. Atoms that constitute phonons  have finite mass, but phonons possess neither intrinsic spin nor charge. Recent studies have defined an effective spin degree of freedom for phonons by introducing the concept of chiral phonons, in which phonons acquire PSAM via the circular motion of atoms about their equilibrium positions [see inset (i) of Fig.~\ref{Scheme}(a)]. The introduction of PSAM naturally raises a deeper question: How are the (pseudo) \textbf{orbital} degrees of freedom defined for quasiparticles such as magnons and phonons? 

To address this question, we recall that both magnons and phonons are collective excitations corresponding, respectively, to spin waves and elastic waves. Their dynamics are governed not only by the microscopic properties of the constituent atoms [e.g.~local spin precession around fixed atomic sites as shown in inset (ii) of Fig.~\ref{Scheme}(a)], but also by the \textbf{emergent} behavior arising from the superposition of wavefunctions with different wavelengths, i.e., the wavepacket dynamics. These wavepacket dynamics are described by the quantum geometry of the Bloch wavefunctions, which means that a wavepacket formalism for the orbital degrees of freedom applies not only to magnons and phonons, but more generally to all fermionic and bosonic particles and quasiparticles in periodic systems \cite{Busch2023,Atencia31122024,To2025}. For instance, in the case of electrons, this quantum geometric effect can give rise to the orbital Hall effect even for $s$-electrons that have zero atomic orbital angular momentum \cite{Busch2023,Pezo2022,Pezo2023}. 

The quantum geometric tensor of a non-degenerate Bloch eigenstate $\vert \psi_{n} \left(\boldsymbol{k} \right) \rangle$ is given by $Q_{ij}^{n}\left(\boldsymbol{k} \right) = \left\langle \partial_{k_{i}}\psi_{n} \left(\boldsymbol{k} \right)\left\vert 1 -P_{n}\left(\boldsymbol{k} \right)\right\vert \partial_{k_{j}}\psi_{n} \left(\boldsymbol{k} \right) \right\rangle$, with $P_{n}\left(\boldsymbol{k}\right) = \vert \psi_{n} \left(\boldsymbol{k} \right) \rangle \langle \psi_{n} \left(\boldsymbol{k} \right) \vert$ the projection operator onto state $\vert \psi_{n} \left(\boldsymbol{k} \right) \rangle$. $Q_{ij}^{n}\left(\boldsymbol{k} \right)$ encodes the curvature [the imaginary part of $Q_{ij}^{n}\left(\boldsymbol{k} \right)$] and the metric [the real part of $Q_{ij}^{n}\left(\boldsymbol{k} \right)$] of a geometric structure. In condensed matter systems these properties are known, respectively, as the Berry curvature and the quantum metric in Hilbert space and can be interpreted as reflecting virtual interband transitions between different Bloch states \cite{Paivi2023}. This implies that interband coupling plays a crucial role in shaping wavepacket dynamics and, consequently, should contribute to emergent physical properties such as the orbital degree of freedom of quantum systems. 

In the case of phonons, when interband coupling is negligible, their dynamics can be well described using a classical picture. In this limit, the total pseudo angular momentum of a phonon can be expressed as $\boldsymbol{J}^{tot}_{ph} = \boldsymbol{J}^{sp}_{ph}+\boldsymbol{J}^{Orb}_{lat}$ where $\boldsymbol{J}^{sp}_{ph} = \sum_{\boldsymbol{R},\alpha}\boldsymbol{u}_{\alpha}\left( \boldsymbol{R} \right) \times \boldsymbol{p}_{\alpha}\left( \boldsymbol{R}\right)$ describes the phonon PSAM and $\boldsymbol{J}^{Orb}_{lat}$ represents the lattice POAM \cite{Park2020}. The POAM, when averaged over a timescale much longer than the phonon period, is given by $\left\langle \boldsymbol{J}^{Orb}_{lat} \right\rangle_{t} = \sum_{\boldsymbol{R},\alpha}\left[\boldsymbol{R} +\boldsymbol{\delta}_{\alpha}\left(\boldsymbol{R} \right) \right]\times M_{\alpha}\frac{d}{dt}\left[\boldsymbol{R} +\boldsymbol{\delta}_{\alpha}\left(\boldsymbol{R} \right) \right]$, where $\boldsymbol{R}$ denotes the position of unit cells; $\boldsymbol{\delta}_{\alpha}\left( \boldsymbol{R}\right)$ is the equilibrium position of atom $\alpha$, which has mass $M_{\alpha}$, within the unit cell at $\boldsymbol{R}$; and $\boldsymbol{u}_{\alpha}\left( \boldsymbol{R} \right)$ and $ \boldsymbol{p}_{\alpha}\left( \boldsymbol{R}\right)$ are, respectively, the displacement from the equilibrium and the momentum of atom $\alpha$. If the equilibrium positions remain fixed, i.e.~there is no rigid-body rotation or translation of the entire system, then the lattice POAM vanishes and only the phonon PSAM can be finite.

In the presence of strong band hybridization, for example in coupled magnon-phonon systems, quantum geometric effects due to interband transitions become significant and can induce nontrivial pseudo angular momentum (PAM) contributions that go beyond the above classical expectations. In such a case the total PAM of a phonon is given by $\boldsymbol{J}^{tot}_{ph} = \boldsymbol{J}^{sp}_{ph}+\boldsymbol{J}^{Orb}_{lat} + \boldsymbol{J}^{Orb}_{QG}$, where $\boldsymbol{J}^{\mathrm{Orb}}_{\mathrm{QG}}$ arises from the quantum geometry of the phonon Bloch wavefunctions. We will show that this term can remain finite even when the lattice orbital contribution $\boldsymbol{J}^{\mathrm{Orb}}_{\mathrm{lat}}$ vanishes. A similar decomposition applies to magnons, with the caveat that magnons do not possess the pseudo spin degree of freedom in the same sense as phonons \cite{Streib2021}. The total pseudo magnon angular moment is thus given by $ \boldsymbol{J}^{tot}_{mag} = \boldsymbol{J}^{Orb}_{lat} + \boldsymbol{J}^{Orb}_{QG}$, which contains only the contributions from global rotation of the lattice ($\boldsymbol{J}^{Orb}_{lat}$) and quantum geometric effects ($\boldsymbol{J}^{Orb}_{QG}$).

We now begin to investigate the orbital degree of freedom in strongly coupled magnon–phonon systems. To treat magnons and phonons on an equal footing, we focus on the OAM, which differs from the conventional POAM by a factor of mass. This distinction is crucial, as magnons, being massless quasiparticles, do not support a classical interpretation of angular momentum. We restrict our analysis to scenarios in which the entire lattice is fixed ($\boldsymbol{J}^{Orb}_{lat} = 0$) and concentrate solely on the contribution from quantum geometry ($\boldsymbol{J}^{Orb}_{QG}$) that is a result of self-rotation of magnon-polaron wavepackets. To make our analysis clearer, we focus on the specific example of magnon–phonon coupling in a 2D antiferromagnetic honeycomb lattice with N\'eel order, in which magnons strongly hybridize with the out-of-plane vibrational (phonon) modes and we can explicitly describe this coupling in the Hamiltonian. While this example allows us to illustrate the important insights and predictions obtained from the quantum geometric perspective on orbital degrees of freedom, we stress that the results are not unique to this example case.  

\textit{Model System} -- The model Hamiltonian of our example system is given by $ H = H_{m} + H_{p} + H_{mp}$ where $H_{m}$ is the Hamiltonian of localized spins whose low-energy excited states are magnons, $H_{p}$ is the phonon Hamiltonian, and $H_{mp}$ is the term describing magnetoelastic coupling and the induced hybridization of magnons and phonons. We consider each of these terms of the Hamiltonian separately:

\begin{enumerate}
    \item $H_m$ follows the anisotropic Heisenberg model given by $ H_{m} = \sum_{i,j} J_{ij}\boldsymbol{S}_{i}\boldsymbol{S}_{j} + \Delta\sum_{i}\left( S_{i}^{z}\right)^{2} + g\mu_{B}B_{z}\sum_{i}S_{i}^{z}$ where  $\boldsymbol{S}_{i}=(S_{i}^{x},S_{i}^{y},S_{i}^{z})$ is the operator of total spin localized at a site $i$ of the lattice. In the first term $J_{ij}$ is the exchange coupling between localized spins at sites $i$ and $j$. In the second term $\Delta$ is the easy-axis anisotropy energy. The sum $\sum_{ij}$ runs over all atom pairs in the lattice up to the third-nearest neighbor. The third term describes the Zeeman interaction, which takes into account coupling to the applied magnetic field $B_{z}$  pointing along the $z$-axis; $g$ is the Land\'{e} $g$-factor; and $\mu_{B}$ is the Bohr magneton. To focus exclusively on the role of magnon-phonon coupling, the Dzyaloshinskii-Moriya interaction (DMI) is omitted from the Heisenberg model.
    
    \item $ H_{p}$ is the effective phonon Hamiltonian given by $ H_{p} = \sum_{i} \frac{\left(p_{i}^{z} \right)^{2}}{2M} + \frac{1}{2} \sum_{ij} u_{i}^{z} \Phi_{i,j}^{z}u_{j}^{z}$. Here $p_{i}^{z}$ and $u_{i}^{z}$ are, respectively, the operators of the out-of-plane momentum and the displacement of the atom at site $i$ of the lattice. $\Phi^{z}$ is a spring constant matrix and M is the mass of the atom. 
    
    \item $H_{mp}$ describes the magnetoelastic coupling and is given by $ H_{mp} = -\xi\sum_{i}\left[ \epsilon^{yz}_{i} \left( S_{i}^{x}S_{i}^{z} + S_{i}^{z}S_{i}^{x} \right) + \epsilon^{xz}_{i} \left( S_{i}^{y}S_{i}^{z} + S_{i}^{z}S_{i}^{y} \right) \right]$ where $\xi$ is the coupling strength and $\epsilon^{xz}_{i}$ and $\epsilon^{yz}_{i}$ are strain functions at the $ith$ site. 
\end{enumerate}
Unless otherwise specified, throughout this letter we employ numerical parameters that describe MnPS$_3$ as a specific example of a 2D AFM with honeycomb lattice and N\'eel order \cite{Bazazzadeh2021}. Specifically, we use:  $J_{1}=0.527$~meV; $J_{2}=0.024$~meV; $J_{3}=0.150$~meV corresponding to the nearest-neighbor, second-nearest-neighbor, and third-nearest-neighbor exchange interactions, respectively; $\Delta=-0.002$~meV; $\xi=0.0292$~meV. For $\Phi^{z}_{i,j}$ we consider only the nearest-neighbor transverse spring-constant matrix, which is diagonal with an elastic constant $\Phi=479.9$~meV/\text{\AA}$^{2}$. These parameters were used to produce the quantitative magnon and phonon dispersion relations shown in Fig.~\ref{Scheme}(b) along the high-symmetry path $K\text{--}\Gamma\text{--}M$. The formation of magnon–polaron states as a result of the coupling term $H_{mp}$ is evidenced by the anti-crossings that emerge where the magnon and phonon branches would intersect in the absence of coupling [see inset of Fig.~\ref{Scheme}(b)]. 

\textit{Orbital degrees of freedom arising from Bloch band quantum geometry} -- 
To derive the OAM arising from self-rotation of a magnon-polaron wavepacket, we consider the wavepacket with momentum $\boldsymbol{k}$ in the nth state, which has eigenenergy $E_{n}(\boldsymbol{k})$ and eigenvector $\vert n(\boldsymbol{k})\rangle$. The self-rotation dynamics of a magnon-polaron wavepacket in this state are described by the orbital moment operator $\hat{\boldsymbol{l}}=\frac{1}{4}\left(\hat{\boldsymbol{r}} \times \hat{\boldsymbol{v}} - \hat{\boldsymbol{v}} \times \hat{\boldsymbol{r}} \right)$. The detailed derivation provided in the Supplemental Material (SM)~\footnote[1]{See Supplemental Material at \url{https://mrsec.udel.edu/publications/}, which includes Ref.~\cite{Huang2021}} shows that the matrix elements of $\hat{\boldsymbol{l}}$ are given by $\boldsymbol{l}_{mn}\left(\boldsymbol{k}\right)=- \frac{i\hbar}{4} \sum_{p\neq m,n}\sigma_{3}^{pp}\textit{N}_{mpn}\left( \boldsymbol{k} \right) \left\langle m\left(\boldsymbol{k} \right)\left\vert \hat{\boldsymbol{v}}_{\boldsymbol{k}}\right\vert p \left( \boldsymbol{k} \right)\right\rangle \times \left\langle p\left(\boldsymbol{k} \right)\left\vert \hat{\boldsymbol{v}}_{\boldsymbol{k}}\right\vert n \left( \boldsymbol{k} \right)\right\rangle$ where $\hat{\boldsymbol{r}}$ and $\hat{\boldsymbol{v}} = -\frac{i}{\hbar} \left[\hat{\boldsymbol{r}},\hat{H} \right]$ are the position operator and the velocity operator, respectively;  $\textit{N}_{mpn}\left( \boldsymbol{k} \right) =\frac{1}{\left[ \boldsymbol{\sigma}_{3}E \left(\boldsymbol{k} \right) \right]_{mm} - \left[ \boldsymbol{\sigma}_{3}E\left(\boldsymbol{k} \right) \right]_{pp}} + \frac{1}{\left[ \boldsymbol{\sigma}_{3}E\left(\boldsymbol{k} \right) \right]_{nn} - \left[ \boldsymbol{\sigma}_{3}E\left(\boldsymbol{k} \right) \right]_{pp}}$ with the ${\bm \sigma}_{3}$ matrix given by ${\bm \sigma}_{3} = \begin{pmatrix} \boldsymbol{1}_{N \times N} & 0 \\ 0 & -\boldsymbol{1}_{N \times N} \end{pmatrix},$ where $\boldsymbol{1}_{N \times N}$ is  $N \times N$ identity matrix and \mbox{$\sigma_{3}^{nn}$} is the $n$th diagonal element of ${\bm \sigma}_{3}$; $E(\boldsymbol{k})$ represents the diagonal matrix whose entries are the eigenvalues of $\hat{H}$. 

When $n \neq m$, $\boldsymbol{l}_{mn}\left(\boldsymbol{k}\right)$ describes the inter-band OAM. When $n = m$ one obtains the intra-band magnon OAM in the Bloch wave of the n$^{th}$ band given by $\boldsymbol{l}_{n}\left(\boldsymbol{k} \right)= -\frac{i\hbar}{2}\sum_{p\neq n}\sigma_{3}^{pp}\frac{ \left\langle n\left(\boldsymbol{k} \right)\left\vert \hat{\boldsymbol{v}}_{\boldsymbol{k}}\right\vert p \left( \boldsymbol{k} \right)\right\rangle \times \left\langle p\left(\boldsymbol{k} \right)\left\vert \hat{\boldsymbol{v}}_{\boldsymbol{k}}\right\vert n \left( \boldsymbol{k} \right)\right\rangle }{ \sigma_{3}^{nn}E_{n}\left(\boldsymbol{k} \right)  - \sigma_{3}^{pp}E_{p}\left(\boldsymbol{k} \right) }$, which corresponds to the imaginary part of the quasi-quantum geometric tensor \cite{Kang2025}. $\boldsymbol{l}_{n}\left(\boldsymbol{k} \right)$ is analogous to the Berry curvature in the conventional quantum geometric tensor and the symmetry constraints that govern the Berry curvature also apply to the intra-band OAM of the magnon-polaron system. Table~\ref{SYM} summarizes the symmetry properties of the intra-band OAM with respect to the wavevector $\boldsymbol{k}$ under time-reversal symmetry (TRS), inversion symmetry (P), and combined parity-time symmetry (P$\oplus$T)~\footnote[2]{The combined symmetry P$\oplus$T, as defined here, refers to the simultaneous presence of both inversion and time-reversal symmetries, and is therefore distinct from the conventional $\mathcal{PT}$ symmetry. In particular, a system can preserve $\mathcal{PT}$ symmetry even if either P or TRS, or even both, are individually broken, whereas $P \oplus T$ requires that both P and TRS independently coexist.}. It is evident that a nonzero intra-band OAM requires the breaking of P$\oplus$T symmetry, which can be achieved via magnetoelastic coupling. 

\begin{table}[ht]
\caption{Symmetry constraints on the intra-band OAM of the magnon-polaron system. }
\begin{tabular}{|c|c|c|c|c|c|ccccccccc}
\hline
\hline
& TRS & P &P$\oplus$T \\
\hline
OAM  &$\boldsymbol{l}_{n}\left(\boldsymbol{k} \right)=-\boldsymbol{l}_{n}\left(-\boldsymbol{k} \right)$  &$\boldsymbol{l}_{n}\left(\boldsymbol{k} \right)=\boldsymbol{l}_{n}\left(-\boldsymbol{k} \right)$ &$\boldsymbol{l}_{n}\left(\boldsymbol{k} \right)=0$   \\
\hline
\hline
\end{tabular}
\label{SYM}
\end{table} 

\begin{figure}
\centering
\includegraphics[width=0.5\textwidth]{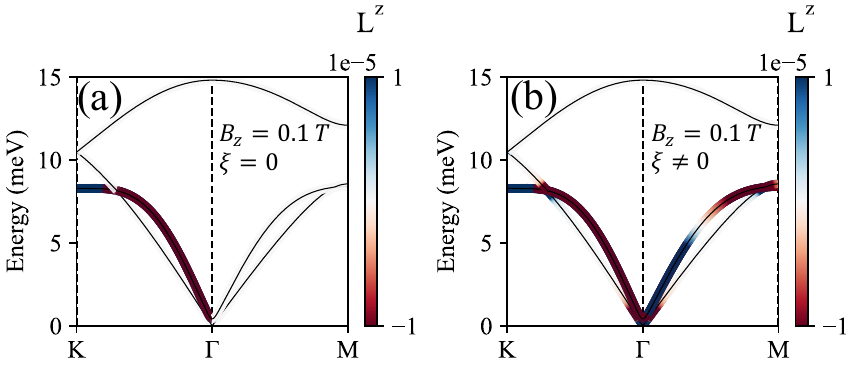}
\caption{Intra-band orbital angular moment-resolved band dispersions of magnon-polarons along the $K - \Gamma - M$ path in a 2D honeycomb antiferromagnet with N\'eel order: (a) without magnon–phonon coupling and (b) with finite magnon–phonon coupling, calculated under an externally applied magnetic field $B_{z}=0.1~\mathrm{T}$.}
\label{ODF}
\end{figure}

To explain this more clearly we consider what happens in the presence and absence of magnetoelastic coupling. In Fig.~\ref{ODF} we present the intra-band orbital angular moment projected band dispersion of magnon-polarons in a 2D honeycomb antiferromagnet  with N\'eel order, plotted along the high-symmetry path $K-\Gamma-M$. When $H_{mp} = 0$ (Fig.~\ref{ODF}a), the magnon band along the $K-\Gamma$ path has a finite OAM due to the breaking of parity-time symmetry. As illustrated in Fig.~S1 of the SM, this finite OAM exhibits odd behavior with respect to the wavevector because of time-reversal symmetry, which remains preserved due to the antiferromagnetic ordering despite the lack of inversion symmetry in the magnetic crystal of a 2D honeycomb AFM with N\'eel order. The OAM vanishes for the magnon band along the $\Gamma-M$ path because this path is P$\oplus$T symmetric in the 2D honeycomb AFM with N\'eel order in the absence of magnon-phonon interaction. The phonon bands do not exhibit any OAM in the absence of magnetoelastic coupling because the phononic crystal on a honeycomb lattice composed of uniform atoms retains TRS, P, and P$\oplus$T symmetry and has no coupling between distinct phonon bands.
 
When $H_{mp} \neq 0$ (Fig.~\ref{ODF}b), the inclusion of magnon-phonon coupling breaks P$\oplus$T symmetry in both the magnonic and phononic subsystems. This symmetry breaking leads to a finite intra-band OAM across the entire magnon-polaron band structure, including along the $\Gamma-M$ path. In other words, through hybridization with magnons, the phonons acquire finite intra-band OAM, which is a manifestation of the transfer of orbital degrees of freedom between the two bosonic quasiparticles. Moreover, when magnetoelastic (ME) coupling breaks TRS the result is an asymmetric OAM distribution such that $\boldsymbol{l}_{n}(\boldsymbol{k}) \neq -\boldsymbol{l}_{n}(-\boldsymbol{k})$, as illustrated in SM Figs.~S2(c,f). In contrast, the inter-band contribution to the OAM (not shown) arises from virtual transitions between different bands and can remain finite even when the intra-band OAM vanishes, provided that there is finite inter-band coupling. This inter-band OAM becomes particularly significant in systems with strong hybridization and narrow band gaps between the coupled bands. 

We now summarize the central conclusion of our study: magnon–phonon coupling enables the conversion of orbital degrees of freedom between distinct bosonic quasiparticles, in this case magnons and phonons, via two distinct mechanisms. First, symmetry breaking induced by the ME interaction allows for the transfer of intra-band OAM between the magnon and phonon and leads to the finite intra-band OAM of the phonon-dominated portion of the magnon-polaron. Second, the ME coupling induces finite inter-band transitions, thereby facilitating the emergence and exchange of inter-band OAM in both magnon- and phonon-dominated portions of the hybridized states. The intra-band OAM corresponds directly to the imaginary part of the quasi–quantum geometric tensor of the bosonic system and the inter-band OAM originates from inter-band transitions and reflects the nontrivial quantum geometry of the Bloch wave functions. This perspective sheds light on the microscopic origin of the ODF in the magnon-polaron system considered here. Specifically, a finite ODF can emerge even in the absence of any rotational orbital angular moment of the lattice due to the intrinsic geometry of the bosonic wave function. This conclusion is universal and extends to both fermionic and bosonic quasiparticles in periodic systems since the quantum-geometric origin of the ODF explored here is determined by the properties of the Bloch wave function and is not restricted to any specific type of particle or quasiparticle. 

\begin{figure}
\centering
\includegraphics[width=0.5\textwidth]{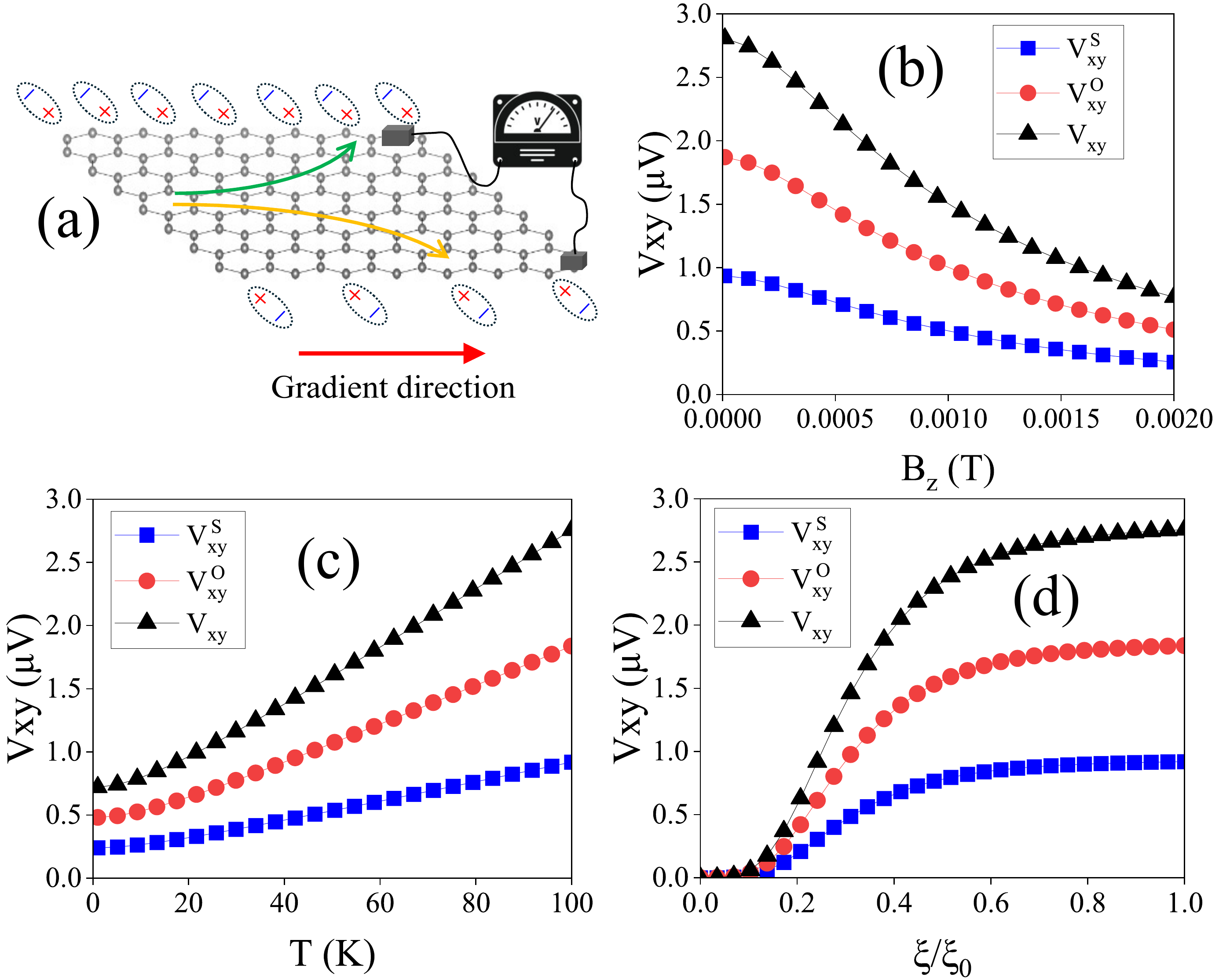}
\caption{(a) Schematic illustration of electric polarization and transverse voltage induced by magnon-polarons accumulating at the sample edges under a longitudinal temperature or strain gradient, which excites magnon-polarons in the system. The resulting transverse voltage along the $y$-direction, including contributions from Berry curvature ($V_{xy}^{S}$) and orbital moment ($V_{xy}^{O}$), is plotted as a function of: (b) the out-of-plane magnetic field $B_z$ at an average temperature of $T = 100$ K; (c) temperature $T$ at zero magnetic field; and (d) Relative magnon-phonon coupling strength ($\xi/\xi_{0}$ with $\xi_{0}=0.0292~meV$) at $T = 100$ K and $B_z = 0$.}
\label{EP_vs_T}
\end{figure}

\textit{Electric response of magnon-polarons as a signature of orbital degree of freedom} -- We now propose an experimental configuration that could be used to probe the orbital degree of freedom in magnon–phonon coupled systems. We begin by noting that the magnon-dominated portion of the magnon-polaron carries a magnetic moment $\boldsymbol{m}$ arising from its intrinsic spin degree of freedom. In the presence of magnon–phonon coupling, this magnetic moment can be partially transferred to the phonon-dominated portion of the magnon-polaron, allowing phonons to acquire a finite magnetic moment via spin–moment interconversion, as illustrated in SM Fig.~S2(a). Consequently, the motion of a magnon-polaron wavepacket can be regarded as the transport of magnetic dipoles. In a classical picture, this motion generates a magnetic dipole current, which, through relativistic effects, induces an electric dipole moment $\boldsymbol{p}=\frac{\boldsymbol{m}\times \boldsymbol{v}}{c^{2}}$ where $\boldsymbol{v}$ is the group velocity of magnon-polaron wavepacket and c is the speed of light. The finite total electric polarization (EP) gives rise to a measurable transverse voltage, $V_{xy}$, along the $y$-direction when a temperature gradient or strain gradient is applied along the $x$-direction to excite the magnon-polaron.  In other words, magnon-polarons carrying opposite OAM and associated electric dipole moments are deflected toward opposite transverse edges, leading to an orbital Hall-like effect. This transverse deflection leads to the accumulation of oppositely oriented electric dipole moments at the edges of the system. When the symmetry between these opposing dipole distributions is broken, a finite transverse voltage $V_{xy}$ emerges, as schematically illustrated in Fig.~\ref{EP_vs_T}(a). The transverse voltage $V_{xy}$ receives two distinct contributions: one from the spin current (via the spin Berry curvature) and the other from the OAM. For clarity, we decompose the total voltage as $V_{xy} = V_{xy}^{S} + V_{xy}^{O}$, where $V_{xy}^{S}$ and $V_{xy}^{O}$ denote the contributions from the spin Berry curvature and the orbital moment, respectively. The precise expression for $V_{xy}$ is provided in the End Matter.

The finite total EP and the resulting transverse voltage can serve as experimental probes of the OAM in the magnon-polaron system, particularly when the OAM contribution dominates over that from the spin Berry curvature. This scenario is realized in MnPS$_3$ under low magnetic fields, as shown in Fig.~\ref{EP_vs_T}(b), where the EP is primarily driven by the magnon-polaron OAM. As reported in Ref.~\cite{To2025}, in the absence of magnon-phonon coupling, but with finite DMI, the OAM and spin Berry curvature contributions differ by only ~20\% in magnitude and have opposite signs, leading to a small net transverse voltage.  In contrast, when magnon-phonon coupling is included, the difference between the two contributions becomes much more pronounced, up to 100\%, with the OAM term becoming approximately twice as large as the spin Berry curvature term and both contributions having the same sign. Notably, the magnitude of the total net EP is enhanced by 5 to 6 orders of magnitude compared to the DMI-only case reported in Ref.~\cite{To2025}. This enhancement arises from the narrow energy gap between two magnon-like bands with phononic character at low magnetic fields, which facilitates strong inter-band transitions. This leads to a large magnon-polaron OAM and spin Berry curvature and consequently a giant polarization. As the magnetic field increases, the band gap widens, suppressing interband transitions and thereby reducing the total EP. 

Overall, the observed electric polarization and the associated transverse voltage, reaching several microvolts at low magnetic fields, provide direct evidence of the ODF arising from the quantum geometry of the magnon-polaron wavefunction. Figures~\ref{EP_vs_T}(c,d) further reveal that both the transverse voltage $V_{xy}$ and its individual contributions from spin and orbital components increase with temperature and the strength of magnon-phonon coupling. Importantly, in the absence of magnon-phonon interaction, the transverse voltage $V_{xy}$ and its associated contributions vanish, as shown in Fig.~\ref{EP_vs_T}(d). In contrast, $V_{xy}$ remains finite even at zero temperature, indicating the persistence of the effect. Further discussion of these features can be found in the SM~\footnotemark[1].  

These results lead to another central conclusion of our study: magnon-phonon coupling significantly facilitates the electric activity of the magnon-polaron system. In particular, it enhances the role of the orbital degree of freedom, making the orbital angular moment the dominant contributor to the magnon-polaron electric response. Additionally, magnon-phonon coupling amplifies both the spin Berry curvature and the magnon-polaron OAM, substantially boosting the system’s electric activity even in the absence of an external magnetic field.  These findings suggest that magnon-phonon coupling provides a promising platform for harnessing the orbital degree of freedom to achieve electrical control over magnetization and phononic transport without relying on external fields.

\textit{Conclusions and perspectives}-- We have presented a theoretical investigation of the orbital degree of freedom in hybrid magnon–phonon systems, unifying the ODF of both magnons and phonons within the framework of orbital angular moment. Focusing on the quantum geometric origin of the OAM in magnon-polaron systems, we demonstrate that magnon–phonon interactions can break the requisite symmetries and couple distinct bands to enable the transfer of intra-band and inter-band OAM between the magnon and phonon contributions to the magnon-polaron. Equally importantly, we demonstrate that magnon–phonon coupling significantly enhances the role of ODF in enabling electrical control over both magnetization and phononic transport. As a concrete example, we show that out-of-plane phonon modes in a honeycomb AFM with N\'eel order can acquire a finite out-of-plane OAM as well as a spin moment via hybridization with magnons. For experimental detection, we propose using the net EP and the resulting transverse voltage at low applied magnetic fields as direct probes of the finite OAM in magnon-polaron systems. 

While this work specifically considers magnetoelastic coupling between magnons and out-of-plane phonon modes, our conclusions regarding the quantum geometric origin of ODF are universal and extend to scenarios involving coupling with in-plane phonon modes. In such cases, phonons may develop a finite PSAM, giving rise to chiral phonons \cite{Ma2024}. These chiral phonons can carry large effective magnetic fields and, through magnon–phonon coupling, exchange PAM with the intrinsic spin of the magnon contribution to a magnon-polaron. As a result, one may expect even stronger net EP in systems where magnons couple to chiral phonons and the OAM of the magnon-polaron becomes entangled with both the magnetic moment of the magnon and the effective magnetic field (PSAM) of the chiral phonon. This interplay opens exciting possibilities, such as electric control and detection of magnon-polarons or efficient light-to-magnon-polaron conversion by exploiting the ODF in magnetic systems. Overall, our work lays the theoretical foundation for the emerging field of phonon orbitronics and establishes a conceptual bridge between phonon and magnon orbitronics, paving the way for a unified framework for harnessing the orbital degree of freedom in condensed matter physics.

\textit{Acknowledgments} -- This research was primarily supported by NSF through the University of Delaware Materials Research Science and Engineering Center, DMR-2011824. We also acknowledge the use of computational resources from the National Energy Research Scientific Computing Center (NERSC), a Department of Energy Office of Science User Facility, through the NERSC award BES-ERCAP 0034471 (m5002).

\bibliography{OAM_PM}

\section*{End Matter} \label{Endmat}
\textit{Expression for the transverse voltage induced by a magnon–polaron wave packet}

Using perturbation theory and assuming the system is in its ground state, we evaluate the energy correction due to a uniform electric field and derive a general expression for the transverse voltage, e.g., along the $y$-direction under a field gradient applied along the $x$-direction, induced by a magnon–polaron in magnetic systems:

\begin{align}
    V_{xy} = V_{xy}^{S} +V_{xy}^{O}
\end{align}  
where
\begin{widetext}
  \begin{equation}
    V_{xy}^{S} = \frac{ g\mu_{B}}{4\pi^{2}\hbar \varepsilon_{0}\chi c^{2}}\sum_{n}\int d\boldsymbol{k}~\left[\Omega_{xy}^{S^{z},n}\left( \boldsymbol{k} \right) - \Omega_{zy}^{S^{x},n}\left( \boldsymbol{k} \right) \right] ln\left\vert e^{-\frac{E_{n,\boldsymbol{k}}}{k_{B}T}}-1\right\vert k_{B}T
\end{equation}
\end{widetext}
and
\begin{widetext}
\begin{align}
    V_{xy}^{O} = &\frac{ g\mu_{B}}{4\pi^{2}\varepsilon_{0}\chi c^{2}}\sum_{n}\int d\boldsymbol{k}~\sigma_{3}^{nn}Im\left[\left\langle n\left(\boldsymbol{k}\right)\left\vert \left(\partial_{k_{y}}\hat{v}_{x} \right)\boldsymbol{\sigma}_{3} \hat{S}^{z}\right\vert n\left(\boldsymbol{k}\right)\right\rangle +4\sum_{q\neq n}\sigma_{3}^{qq}\frac{\left\langle  n\left(\boldsymbol{k}\right)\left\vert \partial_{k_{y}} \hat{H}_{\boldsymbol{k}}\right\vert  q\left(\boldsymbol{k}\right)\right\rangle}{ \left[ \boldsymbol{\sigma}_{3}E_{\boldsymbol{k}} \right]_{nn} - \left[ \boldsymbol{\sigma}_{3}E_{\boldsymbol{k}} \right]_{qq}}\left\langle  q\left(\boldsymbol{k}\right)\left\vert  \hat{j}_{x}^{S^{z}}\right\vert n\left(\boldsymbol{k}\right)\right\rangle  \right]\rho_{n,\boldsymbol{k}} \notag \\
    &- \left(x \leftrightarrow z \right)
\end{align}  
\end{widetext}
Here, $\Omega_{\mu\nu}^{S^{z},n}\left( \boldsymbol{k} \right)$ denotes the $\mu\nu$ component of the spin Berry curvature of the magnon–polaron system, and $\hat{j}^{S^{z}}_{x} = \frac{1}{4} \left(S^{z} \boldsymbol{\sigma}_{3} \hat{v}_{x} + \hat{v}_{x}\boldsymbol{\sigma}_{3} S^{z} \right)$ represents the current operator associated with $S^{z}$ along the $x$-direction. The superscripts $S$ and $O$ label the contributions from the spin Berry curvature and the orbital angular moment of the magnon–polaron wave packet, respectively. These expressions are employed in the main text to analyze the electric activity of the magnon–polaron system, while further quantum-mechanical details on the derivation of the electric polarization and transverse voltage are provided in the Supplemental Material~\footnotemark[1].

\end{document}